\documentclass[aps,amssymb,amsmath,eqsecnum,nofootinbib]{revtex4}
\usepackage{graphics}

\newtheorem{definition}{DEFINITION}[section]
\newtheorem{theorem}{Theorem}[section]

\newtheorem{remark}{Remark}[section]

\newenvironment{hypothesis}{HP: \begin{center}} {\end{center}}
\newenvironment{thesis}{TH: \begin{center}} {\end{center}}
\newenvironment{proof}{\begin{center}PROOF: \end{center}} {$ \blacksquare $}

\begin{document}
 \title{The Aharonov-Anandan phase of a classical dynamical system seen mathematically as a quantum dynamical system}
 \author{Gavriel Segre}
 \email{Gavriel.Segre@msi.vxu.se} \affiliation{International Center
for Mathematical Modelling in Physics and Cognitive Sciences,
University of V\"{a}xj\"{o}, S-35195, Sweden}
 \begin{abstract}
It is shown that the non-adiabatic Hannay's angle of an integrable
non-degenerate classical hamiltonian dynamical system  may be
related to the Aharonov-Anandan phase it develops when it is
looked mathematically as a quantum dynamical system.
\end{abstract}
 \maketitle
 \newpage
 \tableofcontents
 \newpage
 \section{A quantum dynamical system seen mathematically as a classical dynamical system}
Let us start from the following:
\begin{definition}
\end{definition}
\emph{quantum dynamical system:}

 a couple ($ {\mathcal{H}} , \hat{H} $) such that:
 \begin{itemize}
    \item $ {\mathcal{H}} $ is an Hilbert space
    \item $ \hat{H} \in {\mathcal{L}}_{s.a.}( {\mathcal{H}} ) $
    (called the quantum hamiltonian)
 \end{itemize}
where $ {\mathcal{L}}_{s.a.}( {\mathcal{H}} ) $ denotes the set of
all the self-adjoint linear operators over $ {\mathcal{H}} $.

Given a quantum dynamical system  QDS = ($ {\mathcal{H}} , \hat{H}
$) let us introduce the following:
\begin{definition}
\end{definition}
\emph{classical hamiltonian dynamical system associated to QDS}
\begin{equation*}
    CDS [QDS] \; := \; (( {\mathcal{P}} ({\mathcal{H}}) , \omega_{Kahler}[g_{Fubini-Study}])  , H )
\end{equation*}
where:
\begin{itemize}
    \item $ P ({\mathcal{H}}) \; := \; \frac{{\mathcal{H}}}{\sim }
    $ is the projective Hilbert space associated to $
    {\mathcal{H}}$ i.e. the set of equivalence classes in $
    {\mathcal{H}}$ with respect to the following equivalence
    relation:
\begin{equation*}
     |\psi > \sim  | \phi > \; := \: \exists c \in  {\mathbb{C}} \; : \;
    |\psi > = c | \phi >
\end{equation*}
    \item $ \omega_{Kahler}[g_{Fubini-Study}] $ is the K\"{a}hler
    form of the Fubini-Study metric over $  {\mathcal{P}} ({\mathcal{H}}) $ \cite{Nakahara-03}
    \item $ H \in C^{\infty} ( {\mathcal{P}} ({\mathcal{H}} )) $ is the
    hamiltonian defined by:
\begin{equation}
    H ( P_{\psi} ) \; := \; Tr ( \hat{H} P_{\psi} )
\end{equation}
\end{itemize}
 $ P_{\psi} := | \psi > < \psi | $ being the projector associated
 to a $ | \psi > \in {\mathcal{S}} ({\mathcal{H}}) $, where $  {\mathcal{S}}
 ({\mathcal{H}}) \; := \; \{ | \psi > \in {\mathcal{H}} \, : \, < \psi | \psi > = 1
 \}$ is the unit sphere in $ {\mathcal{H}} $ and where we have
 used the fact that:
 \begin{equation}
  {\mathcal{P}} ({\mathcal{H}}) \; \sim_{diff}\; \{ P_{\psi} , | \psi > \in {\mathcal{S}}
  ({\mathcal{H}}) \}
\end{equation}

\smallskip

\begin{remark}
\end{remark}
Since $ ( {\mathcal{P}} ({\mathcal{H}}) , g_{Fubini-Study}) $ is a
K\"{a}hler manifold, its K\"{a}hler form $
\omega_{Kahler}[g_{Fubini-Study}] $ is in particular a symplectic
form, so that  CDS [QDS] is indeed a classical hamiltonian
dynamical system \cite{Marsden-Ratiu-99}.

\smallskip

\begin{remark}
\end{remark}
The classical hamiltonian dynamical system CDS [QDS] has not to be
confused with the classical dynamical system obtained taking the
classical limit of QSD since obviously:
\begin{equation}
    CDS[QDS] \; \neq \lim_{ \hbar \rightarrow 0 } \,
    QDS \; \; \forall \, QDS
\end{equation}
CDS[QDS] is the quantum dynamical system QDS seen mathematically
as a classical hamiltonian dynamical system.

\smallskip

\begin{remark}
\end{remark}
The projective unitary group $ U ( {\mathcal{P}} ({\mathcal{H}}))
\; := \; \frac{U ({\mathcal{H}})}{U(1)} $ of $ {\mathcal{H}} $
acts on $ {\mathcal{P}} ({\mathcal{H}})$ by isometries of $
g_{Fubini-Study} $ that are symplectomorphisms of the symplectic
manifold $ ( {\mathcal{P}} ({\mathcal{H}}) ,
\omega_{Kahler}[g_{Fubini-Study}]) $.

Let $ \rho : G \mapsto  {\mathcal{P}} ({\mathcal{H}})$ be a
projective unitary representation on $ {\mathcal{H}} $ of a Lie
group G. The associated momentum map $ {\mathbf{J}} :
{\mathcal{P}} D_{G} \mapsto L(G)^{\star} $ (where $ L(G)^{\star} $
denotes the dual of the Lie algebra $ L(G) $ of G and where $
{\mathcal{P}} D_{G} $ denotes the essential G-smooth part of
 ${\mathcal{P}} ({\mathcal{H}}) ) $ is equivariant
\cite{Marsden-Ratiu-99}.

Let us consider a  quantum dynamical system QDS = ($ {\mathcal{H}}
, \hat{H} $) such that the associated classical dynamical system
CDS[QDS] is integrable \cite{Arnold-89}.

\begin{remark} \label{rem:Aharonov-Anandan phase related to the non-adiabatic Hannay angle}
\end{remark}
It has been shown in \cite{Marsden-Montgomery-Ratiu-90} that the
Aharonov-Anandan phase (i.e. the non-adiabatic quantum Berry's
phase) of QDS may be related to the non-adiabatic Hannay angle
\cite{Berry-Hannay-88} (i.e. the holonomy of the Hannay-Berry
connection \cite{Montgomery-88},
\cite{Marsden-Montgomery-Ratiu-90}) of CDS[QDS].

\newpage
\section{A classical hamiltonian dynamical system seen mathematically as a quantum dynamical system}
Let us start from the following \cite{Kornfeld-Sinai-00}:
\begin{definition}
\end{definition}
\emph{continuous-time classical dynamical system:}

a couple $  ( ( X \, , \, \sigma \, , \,  \mu ) , \{ T_{t} \}_{t
\in {\mathbb{R}}} ) $ such that:
\begin{itemize}
    \item  $ ( X \, , \, \sigma \, , \,  \mu ) $ is a classical
    probability space
    \item $ \{ T_{t} \}_{t
\in {\mathbb{R}}}  $ is a one-parameter family of automorphisms of
$ ( X \, , \, \sigma \, , \,  \mu ) $, i.e.:
\begin{equation*}
    \mu \circ T_{t}^{-1} \; = \; \mu \; \; \forall t
\in {\mathbb{R}}
\end{equation*}
\end{itemize}

We have seen in the previous section particular instances of the
following notion:
\begin{definition}
\end{definition}
\emph{classical hamiltonian dynamical system}

a couple $ ( ( M , \omega) , H ) $ such that:
\begin{itemize}
    \item $ ( M , \omega ) $ is a symplectic manifold
    \item $ H \in C^{\infty} (M) $
\end{itemize}

One has that:
\begin{theorem}
\end{theorem}

\begin{hypothesis}
\end{hypothesis}

\begin{center}
  $ (( M , \omega ) , H ) $ classical hamiltonian dynamical system
  such that M is compact and orientable
\end{center}

\begin{thesis}
\end{thesis}
\begin{center}
  $ (( M , \omega ) , H ) $ is a continuous-time classical
  dynamical system
\end{center}

\begin{proof}
Let us introduce the classical probability space $ ( M ,
\sigma_{Borel} , \mu_{Liouville} ) $, where $ \sigma_{Borel} $ is
the Borel-$\sigma$-algebra of M and where:
\begin{equation}
  \mu_{Liouville} \; := \;  \frac{ \wedge_{i=1}^{\frac{dim M}{2}} \omega}{ \int_{M}  \wedge_{i=1}^{\frac{dim M}{2}} \omega }
\end{equation}
is the normalized Liouville measure over $  ( M , \sigma_{Borel})
$.

The hamiltonian flow $ \{ T_{t}^{(H)} \}_{t \in {\mathbb{R}}} $
generated by H is a one-parameter family of symplectomorphisms of
$ ( M , \omega ) $ and hence:
\begin{equation}
    \mu_{Liouville} \circ (T_{t}^{(H)})^{- 1} \; = \; \mu_{Liouville} \; \; \forall t
\in {\mathbb{R}}
\end{equation}
\end{proof}

Given a continuous-time classical dynamical system $ CDS := ( ( X
\, , \, \sigma \, , \,  \mu ) , \{ T_{t} \}_{t \in {\mathbb{R}}} )
$ we can adopt Koopman's formalism to introduce the following:
\begin{definition}
\end{definition}
\emph{quantum dynamical system associated to CDS}
\begin{equation}
    QDS[CDS] \; := \; ( {\mathcal{H}} , \hat{H} )
\end{equation}
where:
\begin{itemize}
    \item
\begin{equation*}
  {\mathcal{H}} \; := \; L^{2} ( X , \mu )
\end{equation*}
    \item $ \hat{H} $ is the generator (defined by Stone's Theorem \cite{Reed-Simon-80})  of the strongly-continuous unitary group $ \{ \hat{U}_{t} = \exp( i t \hat{H} ) \}_{t \in
    {\mathbb{R}}}$ such that:
\begin{equation}
     ( \hat{U}_{t} \psi)(x)  \; := \; ( \psi \circ T_{t}) (x) \; \; \psi \in {\mathcal{H}} , t \in
    {\mathbb{R}}
\end{equation}
\end{itemize}

\smallskip

\begin{remark}
\end{remark}
QDS [CDS] has not to be confused with the quantum dynamical system
obtained quantizing CDS since obviously:
\begin{equation}
    \lim_{\hbar \rightarrow 0} QDS [CDS] \; \neq \; CDS \; \;
    \forall \, CDS
\end{equation}
\newpage
\section{Considering the opposite of the remark \ref{rem:Aharonov-Anandan phase related to the non-adiabatic Hannay angle}}

In the remark \ref{rem:Aharonov-Anandan phase related to the
non-adiabatic Hannay angle} we saw that the Aharonov-Anandan phase
of a quantum dynamical system QDS may be related to the
non-adiabatic Hannay angle of CDS[QDS].

In this section we will show that also the opposite occurs, i.e.
that the non-adiabatic Hannay's angle of an integrable classical
hamiltonian dynamical system CDS may be related to the
Aharonov-Anandan phase of QDS[CDS] \footnote{The Aharonov-Anandan
phase of QDS[CDS] was first proposed in \cite{Segre-04} by the
author as the definition of a non adiabatic analogous of Hannay's
angle. At that time I was unaware that non-adiabatic Hannay's
angle was a notion already existing \cite{Berry-Hannay-88}. I
strongly apologize for such an error. In this paper non-adiabatic
Hannay's angle  refers to the notion discovered in
\cite{Berry-Hannay-88} mathematically expressed by the holonomy of
the Hannay-Berry connection \cite{Montgomery-88},
\cite{Marsden-Montgomery-Ratiu-90},
\cite{Chruscinski-Jamiolkowski-04}.}

\smallskip

Given an integrable hamiltonian classical dynamical system $ (( M
, \omega ) , H ) $  with (dim M = 2n)  Liouville's theorem
\cite{Arnold-89} states that a compact and connected level set of
n independent first integrals in involution is diffeomorphic to an
n-dimensional torus
 $T^{n}$ on which the dynamics can be expressed in the action-angle
canonical (i.e. such that $ \omega \; = \; d {\mathbf{I}} \wedge d
{\mathbf{\Phi}} $) variables $ ( {\mathbf{I}} = (I_{1} , \cdots,
I_{n} ) \, , \, {\mathbf{\Phi}} = (\Phi_{1} , \cdots, \Phi_{n} ))
$ as:
\begin{equation}
    \dot{{\mathbf{I}}} \; = \; {\mathbf{0}}
\end{equation}
\begin{equation}
   \dot{{\mathbf{\Phi}}} \; = \; {\mathbf{\Omega}} ({\mathbf{I}})
\end{equation}
where:
\begin{equation}
  {\mathbf{\Omega}} ({\mathbf{I}}) \; := \; \frac{\partial H ({\mathbf{I}})}{\partial {\mathbf{I}} }
\end{equation}
As it is well known there are two cases:
\begin{itemize}
    \item if $ ( {\mathbf{k}} \cdot {\mathbf{\Omega}} :=
     \sum_{i=1}^{n} k_{i} \Omega_{i} \, = \, 0 \; \Rightarrow \; {\mathbf{k}} = {\mathbf{0}} ) \, \, \forall {\mathbf{k}} \in {\mathbb{Z}}^{n}
    $ then the torus $T^{n}$ is said \emph{nonresonant} and the
    dynamics on it is quasi-periodic
    \item if $ ({\mathbf{k}} \cdot {\mathbf{\Omega}}  \, = \, 0 \; \nRightarrow \; {\mathbf{k}} = {\mathbf{0}} ) \, \, \forall {\mathbf{k}} \in
    {\mathbb{Z}}^{n}$ then the torus $ T^{n} $ is said \emph{resonant} and the
    dynamics on it is periodic
\end{itemize}
We will assume that CDS is everywhere non-degenerated, i.e.:
\begin{equation}
 det \frac{\partial {\mathbf{\Omega}} }{ \partial  {\mathbf{I}}}
 \; \neq \; 0
\end{equation}

\smallskip

\begin{remark} \label{rem:on the assumption of globally defined action-angle variables}
\end{remark}
In general the canonical coordinates $ ({\mathbf{I}} ,
{\mathbf{\Phi}}) $ are defined only locally.

This means that considered two different level  sets of the n
independent first-integrals in involution one obtains two
different local charts $ A:= ( U_{A} , \chi_{A} ) $ and $ B := (
U_{B} , \chi_{B} ) $ such that :
\begin{equation}
    \chi_{A} (y) \; = \; ( {\mathbf{I}}_{A} , {\mathbf{\Phi}}_{A}
    )
    \; : \; \omega(y) \, = \, d {\mathbf{I}}_{A} \wedge d
    {\mathbf{\Phi}}_{A} \; \; \forall y \in U_{A}
\end{equation}
\begin{equation}
    \chi_{B} (y) \; = \; ( {\mathbf{I}}_{B} , {\mathbf{\Phi}}_{B}
    ) \; : \; \omega(y) \, = \, d {\mathbf{I}}_{B} \wedge d
    {\mathbf{\Phi}}_{B} \; \; \forall y \in U_{B}
\end{equation}
and where the map $ \psi_{A,B} :  \chi_{B} ( U_{A} \cap U_{B})
\mapsto  \chi_{A} ( U_{A} \cap U_{B}) $:
\begin{equation}
  \psi_{A,B} \; := \;  \chi_{A} \circ \chi_{B}^{-1}
\end{equation}
is infinitely differentiable.

Since the consideration of a symplectic atlas of charts on $ ( M ,
\omega )$ defining a collection of different action-angle
variables simply complicates the situation without adding any
further insight (at least for the matter we are going to discuss)
we will assume that the canonical  action-angle  coordinates $ (
{\mathbf{I}} , {\mathbf{\Phi}}) $ can be extended globally over $
(M , \omega ) $.

\smallskip

 Clearly one has that:
\begin{equation}
    QDS[CDS] \; = \; ( {\mathcal{H}} , \hat{H} )
\end{equation}
where:
\begin{equation}
  {\mathcal{H}} \; = \; L^{2} ( T^{n} , \frac{d {\mathbf{\Phi}}}{(2 \pi)^{n}
  })
\end{equation}
while the strongly continuous unitary group $ \{ \exp ( i \hat{H}
t )\}_{t \in {\mathbb{R}}} $ is specified by its action on the
following basis:
\begin{equation}
    {\mathbb{E}} \; := \; \{ | {\mathbf{n}} > \, := \, \exp ( i
    {\mathbf{n}} \cdot {\mathbf{\Phi}} ) \; , \; {\mathbf{n}} \in
    {\mathbb{Z}}^{n} \}
\end{equation}
given by:
\begin{equation}
   \exp ( i \hat{H}
t ) | {\mathbf{n}} > \; = \; \exp ( i
    {\mathbf{n}} \cdot {\mathbf{\Omega}} t )  | {\mathbf{n}} > \; \; \forall t \in {\mathbb{R}}
\end{equation}

Considered the $U(1)$-principal bundle $ {\mathcal{S}} (
{\mathcal{H}}) ( {\mathcal{P}} ( {\mathcal{H}}) , U(1))  $ it is
well-known that the Aharonov-Anandan geometric phase is given by
the holonomy of the following natural connection one-form $
{\mathcal{A}} \in T^{\star} {\mathcal{P}} ( {\mathcal{H}}) \otimes
L[U(1)]$ (where we denote by L[G] the Lie algebra of a Lie group
G):
\begin{equation}
  {\mathcal{A}}_{\psi} (X) \; := \; i Im < \psi | X > \; \; \psi \in {\mathcal{S}} (
  {\mathcal{H}}) , X \in T_{\psi} {\mathcal{S}} (
  {\mathcal{H}}) \subset {\mathcal{H}}
\end{equation}
A curve $ t \mapsto | \psi(t) > \in {\mathcal{S}} (
  {\mathcal{H}}) $ is horizontal with respect to $ {\mathcal{A}} $
  if and only if:
  \begin{equation}
    < \psi ( t) | \dot{\psi}(t) > \; = \; 0 \; \; \forall t
\end{equation}
So the Aharonov-Anandan geometric phase acquired by QDS[CDS] when
it is subjected to a loop $ \gamma : [ 0 , 1 ] \mapsto
{\mathcal{P}} (
  {\mathcal{H}}) $ such that $ \gamma (0) = \gamma (1) = P_{\psi}
  \, , \, | \psi > \in {\mathcal{S}} (  {\mathcal{H}}) $ is the
  holonomy   $ \tau_{\gamma}^{ {\mathcal{A}}}(  | \psi > )  $.

Let us now consider a family of integrable classical hamiltonian
dynamical systems  $ CDS_{x} \;  := \; ( ( M , \omega ) , H_{x} )
$ where x is a parameter taking values on  a parameters' connected
differentiable manifold P such that $ H_{x} $ depends smoothly by
x and it there exists a point $ x_{0} \in P $ such that $
CDS_{x_{0}} \; = \; CDS$.

Let us then introduce the family of quantum dynamical systems:
\begin{equation}
 QDS[  CDS_{x} ] \; =: \; ( {\mathcal{H}} ,
    \hat{H}_{x} ) \; \; x \in P
\end{equation}

 Let us suppose that the parameter  x   evolves adiabatically
realizing a loop $ \gamma : [ 0 , 1 ] \mapsto P : \gamma(0) =
\gamma(1) = x_{0} $ in P.

The adiabatic limit  under which the  Aharonov-Anandan phase of
QDS[CDS]  reduces to the adiabatic Berry phase of such a quantum
dynamical system may be simply implemented through a suitable
pullback \cite{Bohm-93}.

In the adiabatic limit the basis:
\begin{equation}
  {\mathbb{E}}_{x} \; := \; \{ \,  | {\mathbf{n}} , x  > \; \; {\mathbf{n}}
  \in {\mathbb{Z}}^{n} \, , x \in P  \}
\end{equation}
continues to be formed by eigenvectors of $ \hat{U}_{t} $.

Let us assume that the eigenvalue corresponding to $ |
{\mathbf{n}} , x  > $ is non-degenerate for every $ x \in P $.

Given $ {\mathbf{n}} \in {\mathbb{Z}}^{n} $ let us then introduce
the following map $ f_{{\mathbf{n}}} :  P \, \mapsto \,
{\mathcal{P}} ( {\mathcal{H}}) $:
\begin{equation}
  f_{{\mathbf{n}}} ( x ) \; := \;  P _{| {\mathbf{n}} , x > } \; =
  \; | {\mathbf{n}} , x >  <  {\mathbf{n}} , x |
\end{equation}
Let us then introduce the pullback-bundle $
f_{{\mathbf{n}}}^{\star} {\mathcal{S}} ( {\mathcal{H}}  ) $ of the
U(1)-bundle $ {\mathcal{S}} ( {\mathcal{H}}  ) ( {\mathcal{P}}
({\mathcal{H}}) , U(1) ) $ by $  f_{{\mathbf{n}}} $ and let us
denote by $ f_{{\mathbf{n}}} ^{\star} {\mathcal{A}} $ the
connection on the principal bundle $ f_{{\mathbf{n}}}^{\star}
{\mathcal{S}} ( {\mathcal{H}}  ) $ induced by the connection $
{\mathcal{A}} $ through the pull-back operation; clearly such a
connection is the Berry-Simon connection.

The adiabatic Berry phase developed by QDS[CDS] after the
adiabatic evolution $ \gamma $ is then the holonomy $
\tau_{\gamma}^{f_{{\mathbf{n}}} ^{\star} {{\mathcal{A}}}} $ of the
connection $ f_{{\mathbf{n}}}^{\star} {{\mathcal{A}}} $ along the
loop $ \gamma $.

\smallskip

Let us now consider the Hannay angles of the classical hamiltonian
dynamical system CDS.

At this purpose let us introduce $ E := M \times P $ and the
trivial bundle $ E \stackrel{\pi_{P}}{\rightarrow} P $ where
clearly $ \pi_{P} :  M \times P \mapsto P $ is such that:
\begin{equation}
  \pi_{P} ( y , x) \; := \; x \; \; y \in M , x \in P
\end{equation}
and let us introduce also the other canonical projection $ \pi_{M}
:  M \times P \mapsto P $ defined as:
\begin{equation}
 \pi_{M} ( y , x) \; := \; y \; \; y \in M , x \in P
\end{equation}
Let us observe that the restriction of the pullback $
\pi_{M}^{\star} \omega $ to each fibre $ E_{x} \; := \; \pi_{P}^{-
1}(x) $ is a symplectic form on such a fibre.

Introduced the natural splitting of the total exterior derivative
on $ M \times P $ of a function $ f \in C^{\infty} (  M \times P)
$:
\begin{equation}
    d f \; = \; d_{M} f + d_{P} f
\end{equation}
meaning that, if $ ( y^{1} , \cdots , y^{2n} ) $ are local
coordinates on M and $ ( x^{1} , \cdots , x^{m} ) $ are local
coordinates on P, then:
\begin{equation}
  d_{M} f \; = \; \sum_{i=1}^{2n} \frac{ \partial f}{ \partial
  y^{i}} \, d y^{i}
\end{equation}
\begin{equation}
  d_{P} f \; = \; \sum_{i=1}^{m} \frac{ \partial f}{ \partial
  x^{i}} \, d x^{i}
\end{equation}
let us introduce the following:
\begin{definition}
\end{definition}
\emph{fibrewise hamiltonian vector field $ X_{f} $ corresponding
to f}:
\begin{equation}
    i_{X_{f}} ( \pi_{M}^{\star} \omega ) \; = \; d_{M} f
\end{equation}
Note that $ X_{f}$ is tangent to each fibre $ \pi_{P}^{- 1}(x) $
and hence defines an hamiltonian vector field on $ \pi_{P}^{-
1}(x) $ in the usual sense.

Given a Lie group G:
\begin{definition}
\end{definition}
\emph{family of hamiltonian G-actions on E}

a smooth left action $ \Upsilon : G \times E \mapsto E $ of G on E
such that:
\begin{itemize}
    \item each fibre $ E_{x} $ is invariant under the action
    \item the action, restricted to each fibre  $ E_{x} $, is
    symplectic
    \item it admits a smooth family of momentum maps $
    {\mathbf{J}} : M \times P \mapsto L(G)^{\star} $, i.e., for
    any $ x \in P$, the map $
    {\mathbf{J}}( \cdot ,x) : M  \mapsto L(G)^{\star} $ is a
    momentum map in the usual sense for every $ x \in P$.
\end{itemize}

Given a family $ \Upsilon : G \times E \mapsto E $ of hamiltonian
G-actions on E and an arbitrary tensor T on E let us introduce the
following:
\begin{definition}
\end{definition}
\emph{G-average of T:}
\begin{equation}
    < T > \; := \; \frac{1}{| G |} \int_{G} \Upsilon_{g}^{\star} T \,
    dg
\end{equation}
where $ dg $ is the Haar measure on G and where $ | G | :=
\int_{G} d g $.

Let us now observe that since $ CDS_{x} $ is integrable for every
$ x \in P $  there exists, due to Liouville theorem, a set of
local x-dependent action variables $ {\mathbf{I}}( \cdot \, , \,
x) := ( I_{1}( \cdot \, , \, x) , \cdots , I_{n} ( \cdot \, , \,
x) )$.

For the same reasons exposed in the remark \ref{rem:on the
assumption of globally defined action-angle variables} we will
assume, from here and beyond, that this system is globally defined
on E and, furthermore, that is everywhere non-degenerated, i.e.:
\begin{equation}
 det \frac{\partial {\mathbf{\Omega}} }{ \partial  {\mathbf{I}}}
 \; \neq \; 0
\end{equation}

Let us now look at the n-torus $ T^{n} $ as an abelian Lie group;
we have clearly that:
\begin{equation}
    L( T^{n} )^{\star} \; = \; L( T^{n} ) \; = \; {\mathbb{R}}
\end{equation}

 Under the assumed hypotheses it results defined a family of hamiltonian $ T^{(n)}
$-actions $ \Upsilon : T^{n} \times E \mapsto E $ on E whose
associated smooth family of momentum maps is $ {\mathbf{J}} \; =
\; {\mathbf{I}} : E \mapsto {\mathbb{R}}^{n} $.

\begin{remark}
\end{remark}
Let us observe that chosen at random an initial condition on M
the probability of getting into a resonant torus is zero.

Since the quasi-periodic dynamics on a non-resonant torus is
ergodic, the $ T^{n}$-average and the temporal averages are equal.

\smallskip

Let us introduce the following:
\begin{definition}
\end{definition}
\emph{Hannay-Berry connection on $ E
\stackrel{\pi_{P}}{\rightarrow} P $}

the connection $ {\mathcal{B}} $ on $ E
\stackrel{\pi_{P}}{\rightarrow} P $ such that:
\begin{equation}
    hor_{{\mathcal{B}}}(X) \; := \; < ( 0 , X)  > \; \; \forall X \in T_{x} P
\end{equation}
where $  hor_{{\mathcal{B}}}(X) \in T_{x} P \times T_{y} M $ is
the horizontal lift of a vector $ X \in T_{x} P $ induced by the
connection  $ {\mathcal{B}} $.

Let $ \mu \in {\mathbb{R}}^{n} $ be a regular value of the
momentum map $ {\mathbf{J}} ( \cdot , x ) : M \mapsto
{\mathbb{R}}^{n} $ and let us introduce the following sets:
\begin{equation}
    E_{x}^{\mu} \; := \; {\mathbf{J}}^{-1} ( \mu) \cap \pi_{P}^{-1}(x) \; =
    \; T^{n}
\end{equation}
\begin{equation}
    E^{\mu} \; := \; \cup_{x \in P} E_{x}^{\mu}
\end{equation}
Introducing also the projection:
\begin{equation}
  \pi_{\mu} \; := \; \pi_{P}|_{ E^{\mu}}
\end{equation}
one has that $ E^{\mu} ( P , T^{n}) $ is a torus-bundle over M
\footnote{The first intuitive idea of the fact that the adiabatic
Hannay angle should have been given by the holonomy of a
connection on such a torus-bundle was first proposed in
\cite{Gozzi-Thacker-87}}.

Let us finally introduce the following:
\begin{definition}
\end{definition}
\emph{Hannay-Berry connection on $ E^{\mu} ( P , T^{n}) $}

the restriction of $ {\mathcal{B}} $ to $ E^{\mu} $.

\smallskip

Let us suppose that the parameter  x   evolves  realizing a loop $
\gamma : [ 0 , 1 ] \mapsto P : \gamma(0) = \gamma(1) = x_{0} $ in
P.

The Hannay angle of CDS is then the holonomy $
\tau_{\gamma}^{{\mathcal{B}}}$.

\smallskip

Let us now compare the Aharonov-Anandan phase of QDS[CDS] and the
Hannay angle of CDS.

As we saw the former is the holonomy $ \tau_{\gamma}^{
{\mathcal{A}}}$ over the U(1)-bundle $ {\mathcal{S}} (
{\mathcal{H}}  ) ( {\mathcal{P}} ({\mathcal{H}}) , U(1) )$ while
the latter is the holonomy $ \tau_{\gamma}^{{\mathcal{B}}}$ over
the $ T^{n}$-bundle $ E^{\mu} ( P , T^{n}) $.

Let us first of all make the passage to the Simon's spectral
bundle considering, for each $ {\mathbf{n}} \in {\mathbb{Z}}^{n}
$, the map $ f_{{\mathbf{n}}} : P \, \mapsto \, {\mathcal{P}} (
{\mathcal{H}}) $:
\begin{equation}
  f_{{\mathbf{n}}} ( x ) \; := \;  P _{| {\mathbf{n}} , x > } \; =
  \; | {\mathbf{n}} , x >  <  {\mathbf{n}} , x |
\end{equation}
and taking into account the spectral bundle $
 F := f_{{\mathbf{n}}}^{\star} {\mathcal{S}} ( {\mathcal{H}}  ) $
previously introduced:

such a U(1)-bundle  has the same base space, i.e. P , of the $
T^{n}$-bundle $ E^{\mu} ( P , T^{n}) $ while its fibre $ F_{x} $
in $ x \in P $ is:
\begin{equation}
    F_{x} \; = \; \{ \exp( i \alpha ) | {\mathbf{n}} , x > ,
    \alpha \in {\mathbb{R}} \}
\end{equation}

Given $ {\mathbf{n}} \in {\mathbb{Z}}^{n} $ let us now introduce
the following:
\begin{definition}
\end{definition}
\emph{map of relation between the Hannay angle of CDS and the
Aharonov-Anandan phase of QDS[CDS]:}

the map $ R_{{\mathbf{n}}} : Hol_{{\mathcal{B}}} \mapsto
Hol_{f_{{\mathbf{n}}}^{\star} {\mathcal{A}}  } $:
\begin{equation}
    R_{{\mathbf{n}}} ( \tau_{\gamma}^{{\mathcal{B}}} ) \; = \;  \tau_{\gamma}^{f_{{\mathbf{n}}}^{\star}
    {\mathcal{A}}} \; \; \forall \gamma \in C_{x_{0}} (P)
\end{equation}
where $ Hol_{{\mathcal{B}}} $ is the holonomy group of the
connection $ {\mathcal{B}} $, where $
Hol_{f_{{\mathbf{n}}}^{\star} {\mathcal{A}}  } $ is the holonomy
group of the connection $ f_{{\mathbf{n}}}^{\star} {\mathcal{A}} $
and where:
\begin{equation}
    C_{x_{0}} (P) \; := \; \{ \gamma : [0,1] \mapsto P \, : \, \gamma
    (0) = \gamma(1) \, = \, x_{0} \}
\end{equation}
is the set of loops in P based at $ x_{0} $.

\begin{remark}
\end{remark}
Let us observe that $ R_{{\mathbf{n}}} $ maps the Hannay angle of
CDS into the adiabatic Berry phase of QDS[CDS].

Since the adiabatic Berry phase of QDS[CDS] is a particular case
of the Aharonov-Anandan phase related to it by the pull-back
$f_{{\mathbf{n}}}^{\star}$ we can see R as a map relating the
Hannay angle of CDS and the Aharonov-Anandan phase of QDS[CDS].

\smallskip

The function $ R_{\mathbf{n}} $ maps the holonomy of
${\mathcal{B}} $ associated to a loop $ \gamma $ into the holonomy
of $f_{{\mathbf{n}}}^{\star} {\mathcal{A}}$ associated to the same
loop.

Since $ \tau_{\gamma}^{{\mathcal{B}}} \in T^{n} $ while $
\tau_{\gamma}^{f_{{\mathbf{n}}}^{\star}
    {\mathcal{A}}} \in U(1) $ the map $ R_{{\mathbf{n}}} $ has to be of the form:
\begin{equation}
   R_{{\mathbf{n}}} ( \tau_{\gamma}^{{\mathcal{B}}} ) \; = \; \exp [
   i S ( {\mathbf{n}}  \cdot \tau_{\gamma}^{{\mathcal{B}}})]
\end{equation}
for some $ S: {\mathbb{R}} \mapsto {\mathbb{R}}$.

Considering the case in which $ \bar{\gamma} $ is the constant
loop $ \bar{\gamma}(t) \; := \; x_{0} \; \; \forall t \in [0,1] $
one has that since $ \tau_{\bar{\gamma}}^{{\mathcal{B}}} =
{\mathbb{I}}_{T^{n}} $ and $
\tau_{\bar{\gamma}}^{f_{{\mathbf{n}}}^{\star} {\mathcal{A}}} =
{\mathbb{I}}_{U(1)} $ it follows that:
\begin{equation}
    S(0) \; = \; 0
\end{equation}

\end{document}